\documentclass[%
 reprint,
showpacs,
 amsmath,amssymb,
 aps,
]{revtex4-1}

\usepackage{graphicx}
\usepackage{dcolumn}
\usepackage{bm}

\begin{document}

\preprint{APS/123-QED}

\title{Long ligands reinforce biological adhesion under shear flow} 

\author{Aleksey V. Belyaev}
 \email{aleksey{\_}belyaev@yahoo.com}
 \affiliation{ M. V. Lomonosov Moscow State University, Faculty of Physics, 119991 Moscow, Russia}

\date{\today}

\begin{abstract}
In this work the computer modeling has been used to show that longer ligands allow biological cells (e.g., blood platelets) to withstand stronger flows after their adhesion to solid walls.  Mechanistic model of polymer-mediated ligand-receptor adhesion between a microparticle (cell) and a flat wall has been developed. Theoretical threshold between adherent and non-adherent regimes has been derived analytically and confirmed by simulations. These results lead to a deeper understanding of numerous biophysical processes, e.g., arterial thrombosis, and to the design of new biomimetic colloid-polymer systems.
\end{abstract}

\maketitle


\section{Introduction}
Adhesion is ubiquitous in different fields of physics, material science, physical chemistry and biology. Adhesion mechanisms may be complex and involve various physical interactions, for example, electrostatics, lubrication forces, capillarity, roughness, friction, etc. \cite{Mani, Chu2005, JKR} Unlike non-living matter, the biological cells often rely on ``key-lock'' ligand-receptor binding involving special membrane-anchored proteins - cell adhesion molecules (CAMs) - and the adhesive substrate, e.g. extracellular matrix, collagen, or other another cell \cite{Weikl2009, Rakshit2014, Thomas, HammerLauffenberger1987}. The importance of adhesion in biological systems is supported by numerous examples, such as the formation and growth of bacterial clusters and biofilms \cite{Sircar2013} on medical
implants \cite{chung2007},  adhesion of blood cells to endothelial lining of blood vessels during hemostatic \cite{Heemskerk2009, Furie, Jackson} and immune response \cite{Lawrence91leukocytesroll, Rinker2001}, tissue formation, etc.
The adhesion should be reliable and adjustable, especially when the cells are supposed to function in strong hydrodynamic flows.
One of the most intriguing examples is the adhesion of blood platelets to a damaged blood vessel during hemostatic process \cite{BegentBorn, TurittoBaumgartner, Baumgartner, BelyaevBJ2015, Pivkin, Alber2009, Tossenberger2013, Tossenberger2015, Jackson-2, FurieNature, Stalker2013}. It is provided by the reversible binding between the GPIb-IX-V receptor complex on the platelet membrane and protein ligand - von Willebrand factor (vWf). Normally, vWf exists in a form of long chains - ``multimers'' - that provide platelet aggregation in case of severe bleeding \cite{Jedi, Schneider2007, AlexanderKatz2013, Mistry3480, Kulkarni2000}. In some cases, these aggregates can obstruct blood vessels, limiting blood circulation and causing thrombosis. Understanding the mechanics and regulation of this phenomenon may help in development of anti-thrombotic therapy and lab tests \cite{Diamond2011}, hemocompatible materials for implants \cite{Goodman2005} and design biomimetic colloidal-polymer systems \cite{AlexanderKatz_nature2013}.

Earlier models \cite{Pozrikidis2005-2, ModyKing2013, ModyKing2008-2, luo2016, CHAPMAN19983292, Rinker2001, KrasikHammer2004} and experiments \cite{Kim2010, Coburn2011, Piper1998} underlined the role of reaction rates and loading forces on the dynamics of blood cells subjected to microvascular and arterial flow conditions. Number of leukocyte rolling models  assumed that the ligand length was small in comparison to the cell diameter \cite{HammerApte1992, Reboux2008}. Apparently, this is not the case for vWf-mediated adhesion and aggregation of platelets. One of the essential features of vWf is that these proteins can expand laterally up to several micrometers due to blood flow \cite{Slayter15071985, Novak2070, Schneider2007, Rack2017} and capture widely scattered platelets  from the flowing blood. This protein is also responsible for preventing the bleeding by binding blood platelets to the extracellular matrix or damaged blood vessel walls \cite{Jackson, Coburn2011, Yago}. It is known that vWf deficiency of  concentration in blood causes bleeding disorders, as well as short length of such macromolecular ligands \cite{Denis1998}. At the same time, ultra-long vWf chains may cause thrombotic conditions \cite{Dong4033}.
The role of ligand length in cell adhesion to walls has to be studied and quantified.
The questions are why long tethers are required for efficient blood cell adhesion within the blood flow and how long should these tethers be to accomplish their physiological role. In this work, we try to answer these questions by means of computer simulations.

\begin{figure}
  \includegraphics[width=0.7\columnwidth]{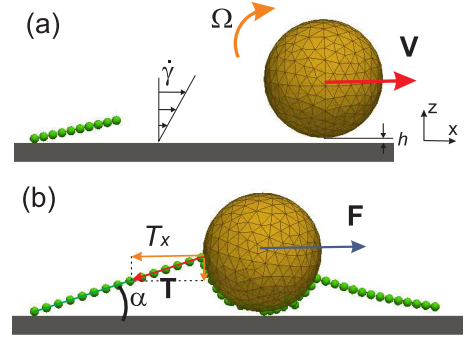}
  \caption{Illustration of  polymer-mediated adhesion mechanics. (a) In the shear flow a spherical particle is rolling and translating along the wall within  distance $h$. The tether can stick to the sphere (b), and if the tether is long enough, it develops a sufficient tension force $T\cos\alpha$ to hold the sphere against the drag force $F$. Yet if $\alpha$ is too big, then the maximum tension $T_{\rm max}$ is reached and the tether detaches (a).}
\label{fig1}
\end{figure}

\begin{figure}
  \includegraphics[width=0.9\columnwidth]{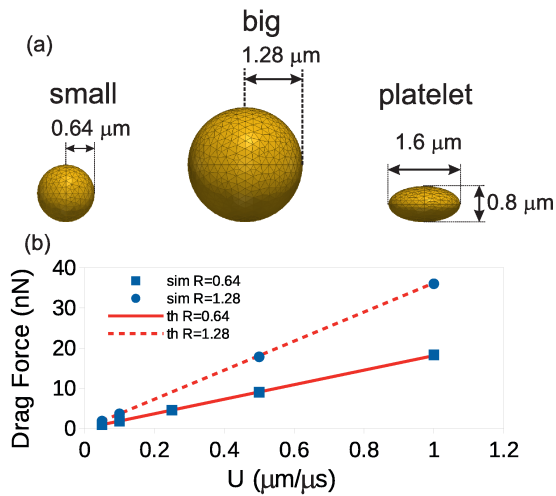}
  \caption{ (a) Three types of particles used in this work. (b) The drag on a sphere held against the flow with constant velocity $U$. Numerical results (squares - for the smaller sphere, circles - for the bigger one) agree with Stokes's law (lines).}
\label{val-sph}
\end{figure}

\section{Methods}
\subsection{Theory}
The adhesion of a microscopic colloid (cell) to a flat wall via binding to a polymer ligand in Couette flow has been studied. A basic mechanistic model has been used for theoretical description. Consider a spherical particle retained against hydrodynamic forces by the polymer near a solid impenetrable wall (Fig.\ref{fig1}). Let us denote the cell radius as $R$ and the tether length as $L$. In case of a long ligand ($R/L\ll 1$), the cell could be represented as a material point. The force balance requires that $F=T \cos\alpha$, where $T$ is the tension force of the tether and $F= A\cdot R^2 \dot\gamma$ is the hydrodynamic drag force with $\dot\gamma$ being the near-wall shear rate. The coefficient $A$ depends on the dynamic viscosity $\mu$ of the fluid, the shape of the particle and the proximity of the walls. According to the Goldman-Cox-Brenner theory \cite{GoldmanCoxBrenner1967}, $A \approx 1.7 \cdot 6\pi\mu$. The tethering polymer can detach from the sphere if $T$ exceeds some threshold value $T_{\rm max}$.  The tethering angle $\alpha$ may be determined on condition that the non-deformed sphere touches the wall, so that $\sin\alpha\approx R/L$. This results in the following expression for the critical shear rate
\begin{equation}
  \label{gamma_cr_1}
  \dot\gamma_{\rm cr} = \frac{T_{\rm max}}{A R^2} \left[1 - (R/L)^2\right],
\end{equation}
which means the adhesion is firm if $\dot\gamma \le \dot\gamma_{\rm cr}$ and impossible otherwise.
In the opposite case ($R/L\gg 1$), the size of the particle becomes substantial, and the corrected formula reads \cite{note1}
\begin{equation}
  \label{gamma_cr_2}
  \dot\gamma_{\rm cr} = \frac{T_{\rm max}}{A R^2} \left[1 - \left(\frac{R}{L+R}\right)^2\right].
\end{equation}
From these expressions we see the advantages of longer ligands: they increase the projection $T_x=T \cos\alpha$ of the tension force $\textbf{T}$ that counterbalances the hydrodynamic drag force $\textbf{F}$  (Fig. \ref{fig1}). This hypothesis has been verified by computer simulations in this work.

\subsection{Computer simulation}

\begin{figure}
  \includegraphics[width=\columnwidth]{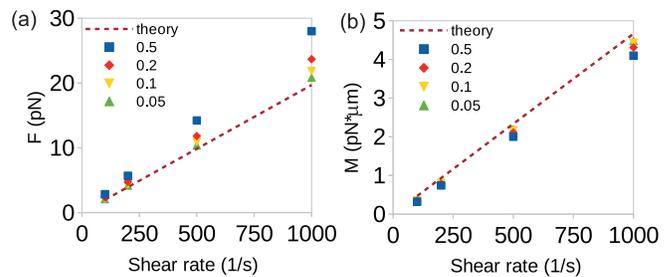}
  \caption{Drag force (left panel) and torque (right) on the immobile sphere ($R=0.64$ $\mu$m) near a wall in shear flow. Symbols correspond to simulation results for different distances $h$ between the sphere and the wall, lines - to the analytical formulas from \cite{GoldmanCoxBrenner1967, ONeill1968}.}
\label{val-sph-wall}
\end{figure}

A 3D computer model has been developed on the basis of the open-source package ESPResSo (version 3.4). A hybrid Lattice Boltzmann (LB) - Particle Dynamics (PD) model has been employed \cite{DunwegLadd, Cimrak2012278, Melchionna2018}. For sake of convenience dimensionless values were used so that $[L]=1$ $\mu$m, $[F]=1$ nN, $[t]=1$ $\mu$s. Lattice Boltzmann BGK-approximation has been used to simulate low Reynolds number hydrodynamics of a viscous fluid in the simulation box \cite{Succi}. Each adhering object (sphere or platelet) immersed into the fluid was represented by a mesh of Lagrangian surface points connected by elastic bonds \cite{Cimrak2012278}. Molecular dynamics (MD) approach has been used to track cell motion over time. The coupling between fluid flow and cell membrane was introduced by a viscous drag force $\mathbf{F}_j=-\xi \Delta\mathbf{u}_j$ exerted on the cell membrane nodes. The hydrodynamic part of the model has been successfully used in prior works \cite{ Cimrak2012278, BelyaevPlos2017}. Before each simulation the ligand polymer (tether) was attached to the bottom wall of the simulation box by one end and stretched along the x-axis. After that, a short equilibration run (500 $\mu$s) was performed before each simulation. The polymer was modeled as a chain of $N$ beads of radius $a=0.05$ $\mu$m connected with elastic elements, for which non-linear FENE potential was used \cite{SIText}.


Spheres of two sizes were used: $R=0.64$ $\mu$m and $1.28$ $\mu$m. The platelet volume was equal to volume of the smaller sphere and had a realistic 2:1 aspect ratio, Fig. \ref{val-sph}(a).
In order to ensure that the model yields correct values of drag forces and torques, validation tests have been performed. A non-deformable sphere was placed into the simulation box in different flow setups. The total drag force $F$ and torque $M$ on the sphere were measured during simulations. The first test was based on a well-known Stokes drag formula: $F_{\rm Stokes} = 6\pi \mu R U$. The sphere was placed into the center of the simulation box and immobilized; a constant velocity condition $(U,0,0)$ was imposed on the boundaries of the box. The results of this test  show a reasonable agreement with Stokes law, Fig.\ref{val-sph}(b). Additionally, a set of validation runs has been performed for the sphere near a flat wall in shear flow, Fig. \ref{val-sph-wall}, and then compared to known theoretical expressions \cite{GoldmanCoxBrenner1967, ONeill1968}. According to these results, the hydrodynamic accuracy of the model is quite high.

\section{Results}

\begin{figure}
  \includegraphics[width=\columnwidth]{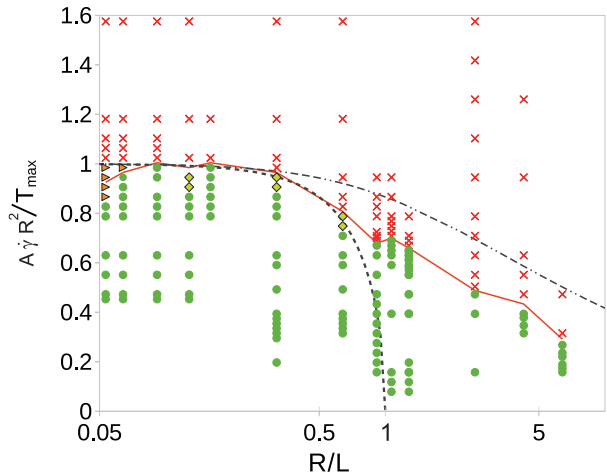}
  \caption{Adhesion regimes diagram: red crosses correspond to cases where the tether was unable to hold the sphere; green circles - where sphere remained attached until the end of simulation; orange triangles - to rolling with stops; light green diamonds - to adhesion when the tethering occurred on the second (or third, etc.) approach. Dashed line corresponds to Eq.({\ref{gamma_cr_1}}) and dash-dotted  line - to Eq.(\ref{gamma_cr_2}). Solid red line is the eye-guide for the boundary between adhesive and non-adhesive regimes.}
\label{res-sph-sphwall-table}
\end{figure}

The first set of results corresponds to the spheres. In the beginning of each simulation the sphere was placed within distance $h_0=2a$ from a wall.
A big set of simulations allowed to plot a map of adhesion regimes (Fig. \ref{res-sph-sphwall-table}). Here, as the theory suggests, the increase of  cell size $R$ relative to the contour length of the tether-polymer ($L=N\cdot r_0$) causes the adhesive bond  to break  at lower shear rates. The simulations support this idea, as seen in Fig. \ref{res-sph-sphwall-table} (see also \textit{Supplementary Movies 1} and \textit{2}). When the ratio is small ($R/L\ll 1$), the threshold of stable adhesion is the highest, as predicted by Eq.(\ref{gamma_cr_1}), and it only slightly depends on $R/L$. For a sphere with $R=0.64$ $\mu$m, the threshold shear rate between the firmly adherent state and the free-flowing regime is approximately 1200 s$^{-1}$. If $R/L\gg 1$, the maximum force that a bond can sustain falls rapidly to zero (in simulations it decreases even faster than predicted by Eq.(\ref{gamma_cr_2})). For ultra-long chains ($N\ge 100$, $R/L <0.06$), an intermediate regime was observed: slow rolling with halts and re-initiation of motion. Apparently, when the force balance is reached, the cell stops. But due to rotation of the sphere around the x-axis, the equilibrium crashes, and the cell starts to roll again (see \textit{Supplementary Movie 3}).

\begin{figure}
  \includegraphics[width=\columnwidth]{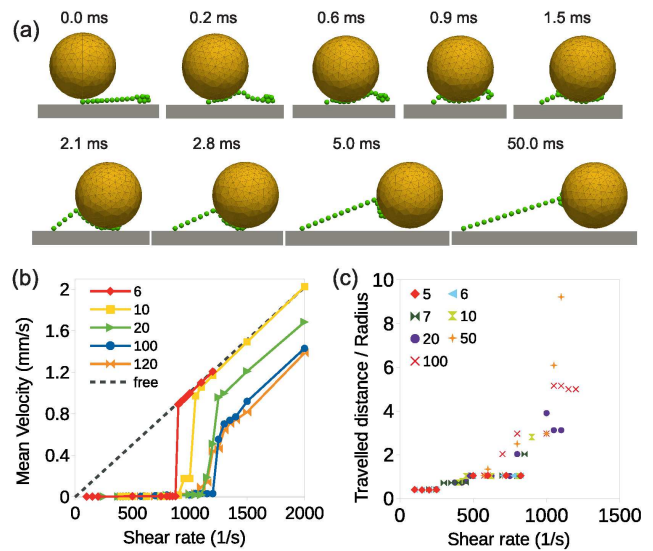}
  \caption{ (a) Series of simulation shots for $R=0.64$ $\mu$m, $N=20$, $\dot\gamma=10^3$ $s^{-1}$. (b) Mean velocity of the sphere ($R=0.64$) plotted against shear rate for different $N$. Dashed line corresponds to the theoretical velocity $V=\dot\gamma(R+h)$ of a free-flowing sphere at distance $h=0.15$ $\mu$m from the wall. (c) Distance traveled by the sphere ($R=0.64$ $\mu$m) before stopping in cases of durable adhesion for different $N$. }
\label{res-sphwall-1}
\end{figure}

\begin{figure}
 \includegraphics[width=\columnwidth]{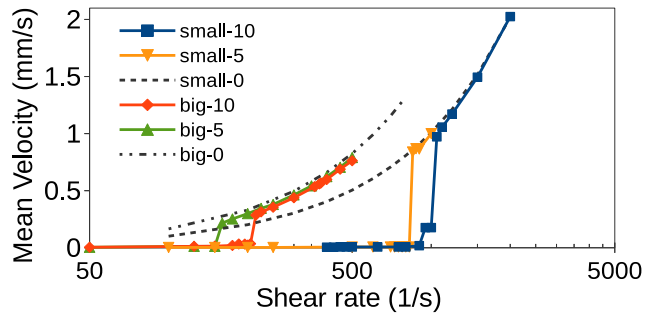}
  \caption{Mean velocity of a sphere plotted against the shear rate for different $N$ and $R$. Lines correspond to theoretical velocity of a free-flowing sphere.}
\label{small-big}
\end{figure}

Mean velocity of the cell was measured during the timespan of 100 ms after the start of each simulation. Simulation snapshots, Fig. \ref{res-sphwall-1}(a), show the initial stages of adhesion. In the beginning, the tension force $T_x$ component (parallel to the wall) is minimal. As the cell moves along the flow, the tethering angle $\alpha$ sharpens. Ideally, $\alpha \rightarrow 0$ for $R/L \rightarrow 0$. Yet, if $T_{\rm max}$ is not sufficient to hold the cell attached to the wall for a given $L=Nr_0$, the tether breaks and the cell starts flowing with the fluid. It was found that longer ligands indeed help to withstand more intensive flows, Fig. \ref{res-sphwall-1}(b). Negligible mean velocity corresponds to lasting adhesion. The transition between adhesion regimes shifts to greater shear rates for longer polymers, in accordance with the theory.
Longer linkers ($> 10$ monomers) also decelerate the cell even in the no-adhesion regime, as there is a period of slow rolling over the polymer chain with consequent detachment. The longer the chain, the longer this period. In cases of stable adhesion the distance traveled by the cell before it stops increases non-linearly with the shear rate, Fig. \ref{res-sphwall-1}(c).

The value of the bond-rupturing shear rate $\dot \gamma_{\rm cr}$ depends strongly on the cell size: if we double the sphere radius, then the drag increases four times (as $F_{\rm drag} \propto \dot\gamma R^2$). Therefore, for the same detachment force  $T_{\rm max}$ and the same $R/L$ (i.e. double the length of ligand and the radius) the critical shear rate for the big cell should be four times smaller than for the small one. This was observed in simulations (Fig. \ref{small-big}): for the small sphere and the 5-monomer-long ligand $\dot\gamma_{cr} \approx 850$ s$^{-1}$, while for the big sphere and the 10-monomer-long ligand $\dot\gamma_{cr} \approx 215$ s$^{-1}$. It should be noted that if the ligand length is kept the same, then $\dot\gamma_{cr}$ decreases even more: for $R=1.28$ $\mu$m and $N=5$ the maximum shear rate is 160 s$^{-1}$. This means that longer polymers are required to hold bigger cells against the stream (see \textit{Supplementary Movie 4}). The blood platelets have a typical size of 1-2 microns and the adhesive ligand (von Willebrand factor) could be 20-50 monomers long (on average), which, as  simulations suggest, is sufficient to withstand drag at wall shear rate of 1000-1200 s$^{-1}$.

\begin{figure}
\includegraphics[width=0.8\columnwidth]{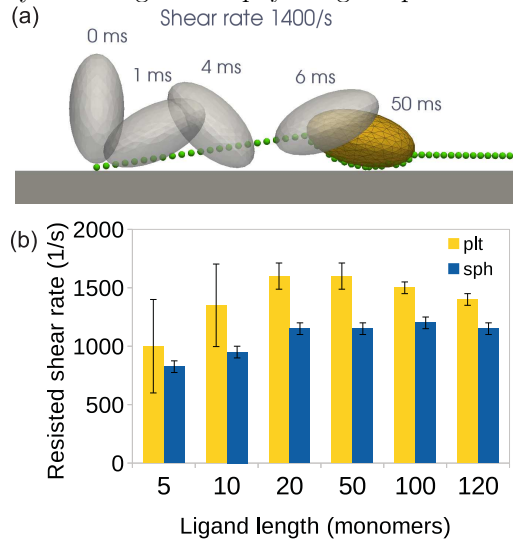}
  \caption{(a) Simulation of the platelet motion before adhering to a wall via binding with grafted von Willebrand factor chain. (b) Diagram comparing the maximum shear rate sustained by the platelet (yellow) and the equivolume sphere (blue). For the platelet, the error bars correspond to deviation due to its initial placement (different orientation of symmetry axes), and for the sphere - to a systematic uncertainty (50 s$^{-1}$) caused by the step-change of shear rate between two consequent simulations. }
\label{res-plt}
\end{figure}

Blood platelets are not spherical: normally they are oblate spheroids. In order to investigate the effect of the cell shape, simulations with a platelet-shaped particle have also been performed in a range of shear rates. Different ligand lengths and different initial placements of the platelet were used.
It has been discovered that platelets demonstrate a more complicated motion than spheres \cite{PozrikidisJFM, ModyKing2013}. Due to the shape and the non-penetration condition, the center of mass of the platelet jumps and falls as the cell rolls over the adhesive wall Fig. \ref{res-plt}(a), in unlike the sphere.  The results of the simulations, presented in Fig.\ref{res-plt}(b), suggest that platelets can sustain greater wall shear rates than spheres. This is due to their streamlined shape, lower profile and, consequently, weaker drag force (see \textit{Supplementary Movie 5}).

\section{Conclusion}

In conclusion, this study shows that longer tethers are preferable for binding of blood platelets to injury, thrombus or extracellular matrix due to the more favorable direction of the tether tension force. The oblate shape and small size allow platelets to minimize the drag and promote aggregation. Hopefully, these findings will deepen the physical insight into physiological phenomena that rely on cell adhesion, e.g., thrombosis and hemostasis.

\begin{acknowledgments}
 This research was supported by the Russian Science Foundation (project number 17-71-10150). 
 The simulations were carried out using computational resources of the Supercomputing Center of 
 M. V. Lomonosov Moscow State University (Lomonosov and Lomonosov-2) \cite{Lomonosov}.
\end{acknowledgments}

\bibliography{pltvwf} 

\begin{thebibliography}{61}
\bibitem [{not()}]{note1}%
  \BibitemOpen
  \href@noop {} {}\bibinfo {note} {Strictly speaking, the rotation of the cell
  should be taken into account from the balance of torques. But for spherical
  cells it only lifts the point of polymer-to-cell binding to a distance
  $\approx R$ from the wall, as observed in simulations.}\BibitemShut {Stop}%
\end{thebibliography}

\end{document}


\title{\textbf{Supplementary information}. \\ \textit{Long ligands reinforce biological adhesion under shear flow}}

\author{Aleksey V. Belyaev}
 \email{aleksey{\_}belyaev@yahoo.com}
 \affiliation{ M.V. Lomonosov Moscow State University, Faculty of Physics, 119991 Moscow, Russia}

\maketitle

\section{Computer model details}

A 3D computer model has been developed on the basis of the open-source package ESPResSo (version 3.4). A hybrid Lattice Boltzmann (LB) - Particle Dynamics (PD) model has been employed \cite{DunwegLadd, Cimrak2012278, Melchionna2018}. For the sake of convenience, dimensionless values were used for convenience so that $[L]=1$ $\mu$m, $[F]=1$ nN, $[t]=1$ $\mu$s. Bhatnagar-Gross-Krook approximation \cite{Succi} was used to simulate low-Reynolds number shear flow of a fluid with density $\rho=1.0$ and kinematic viscosity $\nu=1.5$. In most of the presented simulations, a lattice unit for LB was set to $\Delta x = [L]$, as no significant change of the numerical results was found after the refinement of the lattice.

A $16\times8\times8$ $\mu$m$^3$ simulation box was confined between two impermeable walls, with constant velocity maintained on the upper wall to create a linear shear (Couette) flow in the system. The no-slip condition was introduced on the bottom wall via a half-way bounce-back rule \cite{DunwegLadd,Succi}. For long ligands ($>50$ monomers), a longer box (32$\times$8$\times$8 $\mu$m$^3$) was used. Periodic boundary conditions were imposed in x and y directions.

 Each cell is represented by a mesh of Lagrangian surface points connected by elastic bonds \cite{Cimrak2012278}: 304 mesh nodes for the small sphere and the platelet, and 727 nodes for the big sphere. Molecular dynamics (MD) approach has been used to track cell motion in time. Dual time-step scheme was used: in each LB time step, 100 MD steps were performed. The cells were assumed non-deformable provided by the existence of a cytoskeleton in free-flowing blood platelets \cite{Jackson}, thus were the membrane elastic constants have been chosen: $k_s= 1.0$ for stretching, $k_b=1.0$ for bending, $k_{a}=1.0$ and $k_v=1.0$ for area and volume conservation respectively \cite{BelyaevPlos2017}.
 The coupling between fluid flow and cell membrane has been introduced by a viscous drag force $\mathbf{F}_j=-\xi \Delta\mathbf{u}_j$ exerted on the cell membrane nodes.  The coupling parameter was chosen to be $\xi=0.125$ (in model units) for the small sphere and the platelet, and $0.15$ for the big sphere.

 A very short-range empirical repulsive potential between the wall and the cell membrane points was introduced \cite{ModyKing2008-1} in order to reproduce joint effect of wall (or glycocalyx) elasticity and electrostatic repulsion
\begin{equation}
    U_{\rm rep}(z) = 0.0001\cdot(z+0.2)^{-1.2}
\end{equation}
with $z$ being the distance from the wall, and the cut-off distance  $r_{\rm cut}=0.25$ model units.

The ligand polymer (or tether) was modeled as a chain of $N$ beads of radius $a=0.05$ model units connected with springs; $N$ was varied from 2 to 120. The chain was attached to the bottom surface by one end and stretched along the x-axis. After that a short equilibration run (500 $\mu$s) was performed before each simulation.  Finitely Extensible Non-linear Elastic (FENE) potential was used for springs
\begin{equation}
  U_{\rm FENE}(r) =-0.5 K \Delta r_{\rm m}^2 \ln\left[1-(\Delta r/\Delta r_{\rm m})^2\right],
\end{equation}
where $\Delta r = r-r_0$, $r_0= 2a$ is the resting length of the bond between monomers, $\Delta r_{\rm m}=1.0$ is the maximal bond stretch, $K=200$ $k_BT/a^2$ is the polymer stiffness, $k_BT = 4\times 10^{-6}$.

The repulsion between monomers has been introduced to prevent their overlapping. Furthermore, a small hydrophobic attraction between the monomers has been introduced. In general, interaction between monomers in simulations was governed by Lennard-Jones (truncated and shifted) potential
\begin{equation}
  U_{\rm mono-mono}(r) = 4\epsilon \left[ \left(\frac{\sigma}{r}\right)^{12} - \left(\frac{\sigma}{r}\right)^{6}\right]
\end{equation}
with $\sigma= r_0/1.122$, $\epsilon = 0.2$ $k_BT$ and cut-off distance $r_{\rm cut}=2r_0$.

Morse potential (truncated and shifted) was used to describe adhesive interactions between the polymer and the cell
\begin{equation}
  U_{\rm adh}(r)=U_0 \left\{ \exp[-200(r-a)] -2 \exp[-100(r-a)] \right\}
\end{equation}
with cut-off distance equal to $3a$. The amplitude of the adhesive potential was chosen as $U_0=0.0005$, which determined the tether tension of bond rupture $T_{\rm max}\approx 0.0025$ nN. The latter value was measured in a number of independent pulling simulations with a sphere and a 2-monomer-long linker, and it complies with experimental data \cite{Kim2010}.

\section{Captions for Supplementary Movies}

1. \textbf{SMovie1}. The effect of ligand length on polymer-mediated adhesion of a spherical cell ($R=0.64$ $\mu$m) to a flat wall in shear flow. The short ligand (6 monomers) cannot hold the cell, while the long one provides tethering. Shear rate $\dot\gamma=1000$ s$^{-1}$. The simulations have been performed independently (in order to exclude hydrodynamic interference) and overlapped for presentation purposes only.

2. \textbf{SMovie2}. Polymer-mediated adhesion of a spherical cell ($R=0.64$ $\mu$m) to a flat wall in shear flow at 1000 s$^{-1}$ and 1400 s$^{-1}$.

3. \textbf{SMovie3}. The effect of cell size on polymer-mediated adhesion of spheres. The same 10-monomer ligand has been used in both simulation runs. The shear (Couette) was directed along the x-axis. During the simulation, the smaller sphere ($R=0.64$ $\mu$m) was immobilized by a  tether, while  the bigger sphere ($R=1.28$ $\mu$m)  continued to roll. One can see that the bigger cell is pulled to the wall on its second approach to the polymer (due to periodicity of the simulation box), but the tether is too short to withstand the hydrodynamic drag.  The simulations were performed independently (in order to exclude hydrodynamic interference) and overlapped  for presentation purposes only.

4. \textbf{SMovie4}. The ``rolling-with-stops'' regime observed for the longer ligands ($N=120$) at shear rates 1150 s$^{-1}$ and 1200 s$^{-1}$. One can see that after the sphere stops, it rotates around its axis of symmetry, parallel to the x-axis of the coordinate system. This rotation eventually agitates the force balance and the sphere starts rolling again. For 1200 s$^{-1}$ stoppage periods were much shorter than for 1150 s$^{-1}$.

5. \textbf{SMovie5}. Computer simulations of von Willebrand factor-mediated adhesion of blood platelets to the wall. (A) Side view of the platelet motion for two  shear rates: for 1400 s$^{-1}$ the lasting adhesion has been observed, but not for 2000 s$^{-1}$. The platelet performs the flipping motion due to its shape and is also  repelled from the wall due to geometrical constraints. Spheres, on the other hand, maintain quasi-steady motion.  (B) The oblate shape, however, appears to be more favorable for  adhesion than the sphere due to its lower hydrodynamic profile. This is a comparison of  simulations for the platelet adhering to the 100-monomer-long ligand (vWf) in 1400 s$^{-1}$ shear flow and the simulation for the equivolume sphere under the same hydrodynamic conditions. The platelets bind regardless of initial orientation, while the sphere rolls and detaches from the tether. For each cell, simulations have been performed independently (in order to exclude hydrodynamic interference) and are overlapped only for the presentation purposes only. The arrows indicate the direction of the fluid flow.

\bibliography{pltvwf}